\documentclass[a4paper,10pt,twoside]{cpc-hepnp}
\usepackage{CJK,upgreek,fancyhdr}
\usepackage{multicol}
\usepackage{graphicx}
\usepackage{booktabs}
\usepackage{amssymb,bm,mathrsfs,bbm,amscd}
\usepackage[tbtags]{amsmath}
\usepackage{lastpage}

\begin{document}
\begin{CJK*}{GB}{gbsn}

\fancyhead[c]{\small Chinese Physics C~~~Vol. 41, No. 2 (2017) 026002}
\fancyfoot[C]{\small 010201-\thepage}


\title{
  A method to monitor and measure the water transparency in LHAASO-WCDA using cosmic muon signals
\thanks{Supported by U1332201, U1532258 and NSFC (No.11375224 and 11675187)}}

\author{%
\quad Hui-Cai Li$^{1}$ \email{lihuicai@ihep.ac.cn}%
\quad Zhi-Guo Yao$^{2}$
\quad Chun-Xu Yu$^{1}$
\quad Ming-Jun Chen$^{2}$
\\\quad Han-Rong Wu$^{2}$
\quad Min Zha$^{2}$
\quad Bo Gao$^{2}$
\quad Xiao-Jie Wang$^{2}$
\\\quad Jin-Yan Liu$^{1}$
\quad Wen-Ying Liao$^{1}$ £¨for the LHAASO collaboration£©
}

\maketitle

\address{%
$^1$ Nankai University, Tianjin 300071, China\\
$^2$ Institute of High Energy Physics, Chinese Academy of Sciences, Beijing 100049, China\\
}

\begin{abstract}
The Large High Altitude Air Shower Observatory (LHAASO) is to be built at Daocheng, Sichuan Province, China. As one of the major components of the LHAASO project, a Water Cherenkov Detector Array (WCDA), with an area of 78,000~$\rm m^{2}$, contains 350,000~tons of purified water. The water transparency and its stability are critical for successful long-term operation of this project. To gain full knowledge of the water Cherenkov technique and investigate the engineering issues, a 9-cell detector array has been built at the Yangbajing site, Tibet, China. With the help of the distribution of single cosmic muon signals, the monitoring and measurement of water transparency are studied. The results show that a precision of several percent can be obtained for the attenuation length measurement, which satisfies the requirements of the experiment. In the near future, this method could be applied to the LHAASO-WCDA project.
\end{abstract}



\footnotetext[0]{\hspace*{-3mm}\raisebox{0.3ex}{$\scriptstyle\copyright$}2013 Chinese Physical Society and the Institute of High Energy Physics of the Chinese Academy of Sciences and the Institute of Modern Physics of the Chinese Academy of Sciences and IOP Publishing Ltd}%

\begin{multicols}{2}

\section{Introduction}\label{sect:introduction}

In very-high-energy gamma ray astronomy, the water Cherenkov technique has the unique advantage of a much better background rejection power than other ground particle detectors such as plastic scintillators and RPCs. This has already been well demonstrated by simulations, and in practice by the Milagro experiment. New generation facilities, like HAWC~\cite{DeYoung2012} and LHAASO~\cite{cao2014}, that adopt this technique and have a larger area will be able to achieve a sensitivity of more than an order of magnitude better than current experiments.

As a major component of LHAASO, the water Cherenkov detector array (WCDA), with an area of 78,000~$\rm m^2$ and 350,000~tons of purified water, is planned to be built in a couple of years at Mount Haizi, Daocheng, at an altitude of 4410~m a.s.l. The main purpose of the WCDA is to survey the northern sky for VHE gamma ray sources. The detector efficiency is a crucial factor for obtaining high sensitivity, real-time monitoring and accurate calibration of the detector. It is also of importance for achieving good spectrum measurement of the targeted gamma ray sources.

The whole WCDA consists of 3 large tanks, two of which have an area of 150~m$\times$150~m, and the other an area of 300~m$\times$110~m (Fig.~\ref{fig:wcdalayout}). Each tank is subdivided into cells with an area of 5~m$\times$5~m, separated by black plastic curtains to prevent the cross-talk of light between cells. In each cell, there are 1 or 2 PMTs deployed at the bottom, facing upwards with an effective water depth of 4~m above the photo-cathode. When a shower particle goes through a detector cell, it will interact with the water and consequently yield some Cherenkov light, some of which may arrive at the photo-cathode of the PMTs. This then produces electric signals, including time and charge information, that are useful for the reconstruction of the shower. Since the Cherenkov photons have to travel a certain distance before hitting the PMTs, the absorption or scattering of the photons by the water molecules or impurities in the water plays an important role.

The water transparency and its stability are critical for successful long-term operation of WCDA. Given the effects of bacteria, dust and ions, the absorption length of natural water from such as wells or streams is usually only several meters (e.g., 6~m). Then the loss of the Cherenkov light generated by the air-shower secondaries in the water is very large, about 50\%. Therefore, the water in the tank has to be purified to a particular level, and recirculated at some rate of flow to kill bacteria and to destroy the organic carbon that the bacteria produce and live with. When the experimental experience of the WCDA prototype array is taken into account, the attenuation length is set larger than 15~m. With an assumption that water attenuation length can be changed from 15~m to 40~m among all WCDA detector cells at different running conditions, nearly 18\% detector efficiency difference can be found. Therefore, it is very important to accurately measure and monitor the water transparency (attenuation length) and its variation, as well as its uniformity in the tank, with some dedicated instruments and analysis methods.

\begin{center}
  \centering\includegraphics[width=0.7\linewidth]{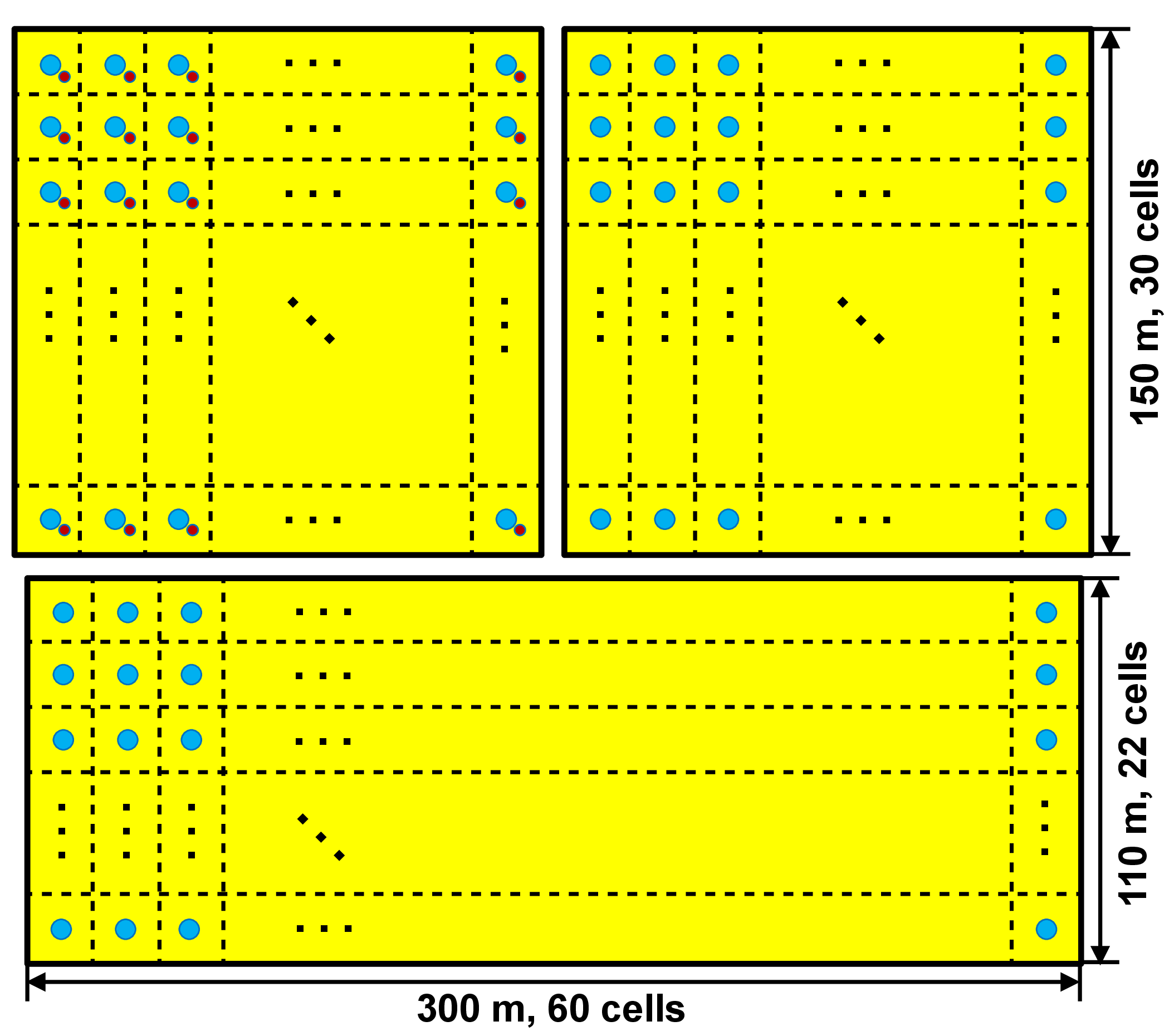}
  \figcaption{Schematic of the WCDA layout.}
  \label{fig:wcdalayout}
\end{center}

\begin{center}
  \centering \includegraphics[width=0.6\linewidth]{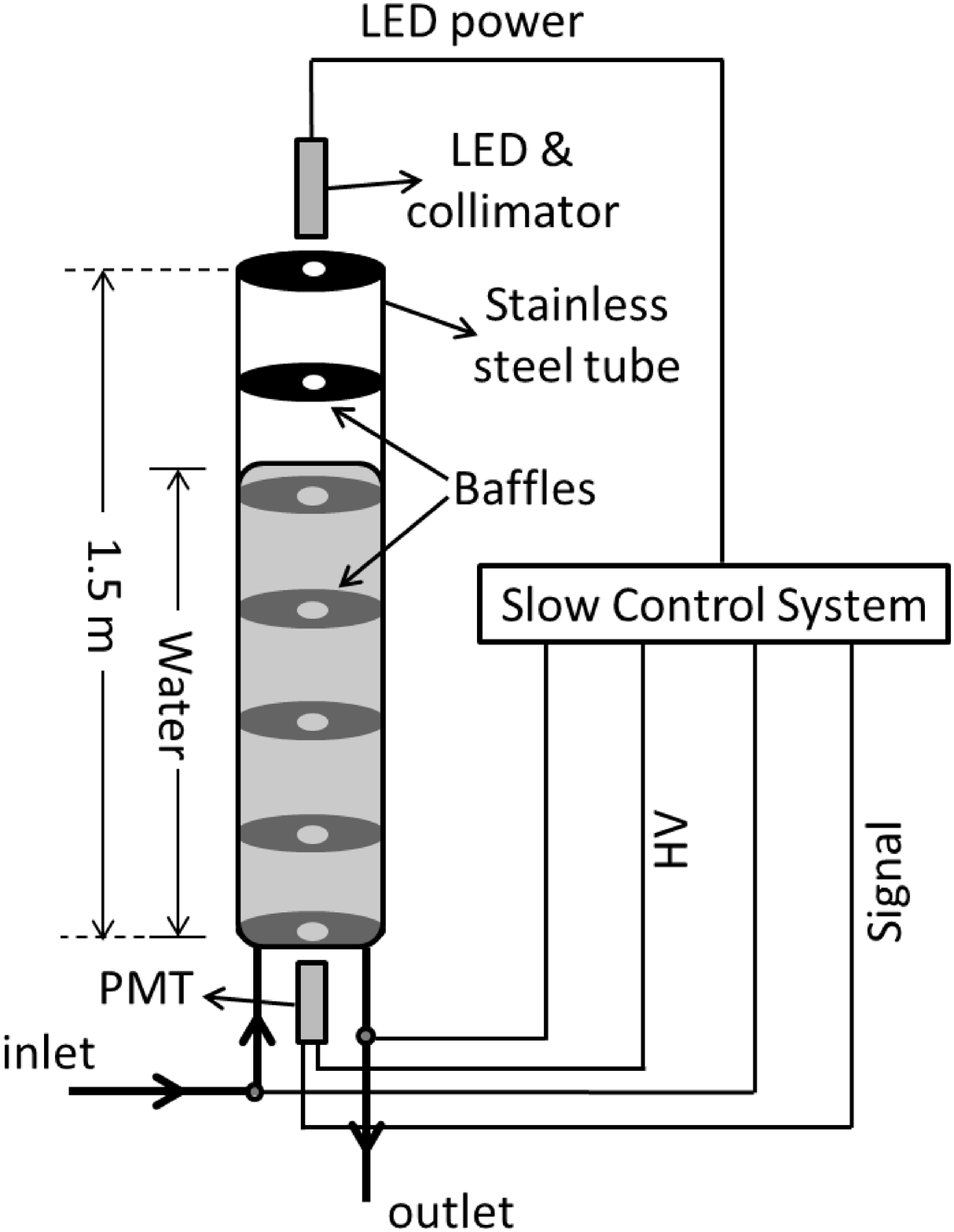}
  \figcaption{A schematic of the device that performs the attenuation length measurement~\cite{an2013}.}
  \label{fig:tubedevice}
\end{center}

The usual method to measure the water transparency is to use a tube-like device, shown in Fig.~\ref{fig:tubedevice}. The tube, standing vertically, is a container of the water sample being measured. Some baffles are assembled in the tube, for the purpose of blocking the scattered light. An
LED with wavelength around 405~nm is fixed at one top of the tube, emitting  pulsed photons through two tiny holes, so only collimated light can pass through the tube. A PMT is installed at the other end of the tube to collect the photons, and the charge of the converted electric signals is digitized with a dedicated electronic system. At the beginning of every measurement, the tube is filled to maximum capacity with the water sample. Then the water is drained several times to adjust the water depth in the tube. At each different water depth, ten thousand signals of the LED pulsed light produced with the same intensity are collected. Finally, the attenuation length of the water can be derived by an exponential law fitting to the collected light intensity as a function of the water depth in the tube.

In the WCDA experiment, we plan to deploy several such  devices, which could be operated continuously to monitor the water transparency at different places in the tanks. Basically, the water transparency of the whole tank can be monitored after long term operation. However, on some occasions, if the water transparency changes rapidly and so many cells need to be measured, this solution is not very convenient and the maintenance cost could be huge. Therefore, a new and more efficient method to monitor and measure the water transparency is necessary.

Cosmic muons, whose flux is estimated to be 300~Hz/$\rm m^2$ at the altitude of 4400~m a.s.l., produce slightly different signals in the PMT of a cell from the electromagnetic components. Most muons can pass through the water, yielding large signals, and the intensity of signals is closely related to the geometry of muon-track. Combined with the simulation and the data of the WCDA prototype array, a method to measure the water transparency can be obtained after detailed analysis.

In Section~2, the WCDA prototype array and its water purification and recirculating system, as well as the water transparency measured with a tube device in a certain period, are introduced. In Section~3, the features of the single-channel signals from the data are described and explained. In Section~4, the analysis of the second peak in the spectrum of the single channel signals and its correlation with the water attenuation length is presented. Finally, the study is summarized and concluded in Section~5.

\section{WCDA prototype array}\label{sect:ptarray}
 To gain full knowledge of the WCDA detector, a prototype array was built at Yangbajing in 2010 and has already been operating for about three years.

\subsection{Water tank}
The prototype array~\cite{an2013} of the water Cherenkov detector is located around 15~m northwest of the ARGO-YBJ experiment hall. A schematic of the tank is shown in Fig.~\ref{fig:pond}. The effective dimensions of the tank are $15~\rm m\times 15~\rm m$ at the bottom, with the tank wall concreted upward along a slope of 45 degrees until it is $5~\rm m$ in height, so it is $25~\rm m \times 25~\rm m$ at the top. The whole tank is partitioned by black curtains into $3 \times 3$ cells, each of which is $5~\rm m \times 5~\rm m$ in size. A PMT is deployed at the bottom-center of each cell, facing upwards to collect the Cherenkov photons generated by air shower particles in water. The size of the black curtains is $4.5~\rm m \times 3~\rm m$, so the light isolation between cells is not completely tight, and some Cherenkov photons from nearby cells can be collected. In the array, two kinds of 8-inch PMTs are deployed, Hamamatsu R5912 and ET 9354KB. To control the quality of the water, a facility for purifying the water was built near the tank, recirculating the water via the pipes installed in the tank.

\begin{center}
  \centering \includegraphics[width=0.9\linewidth]{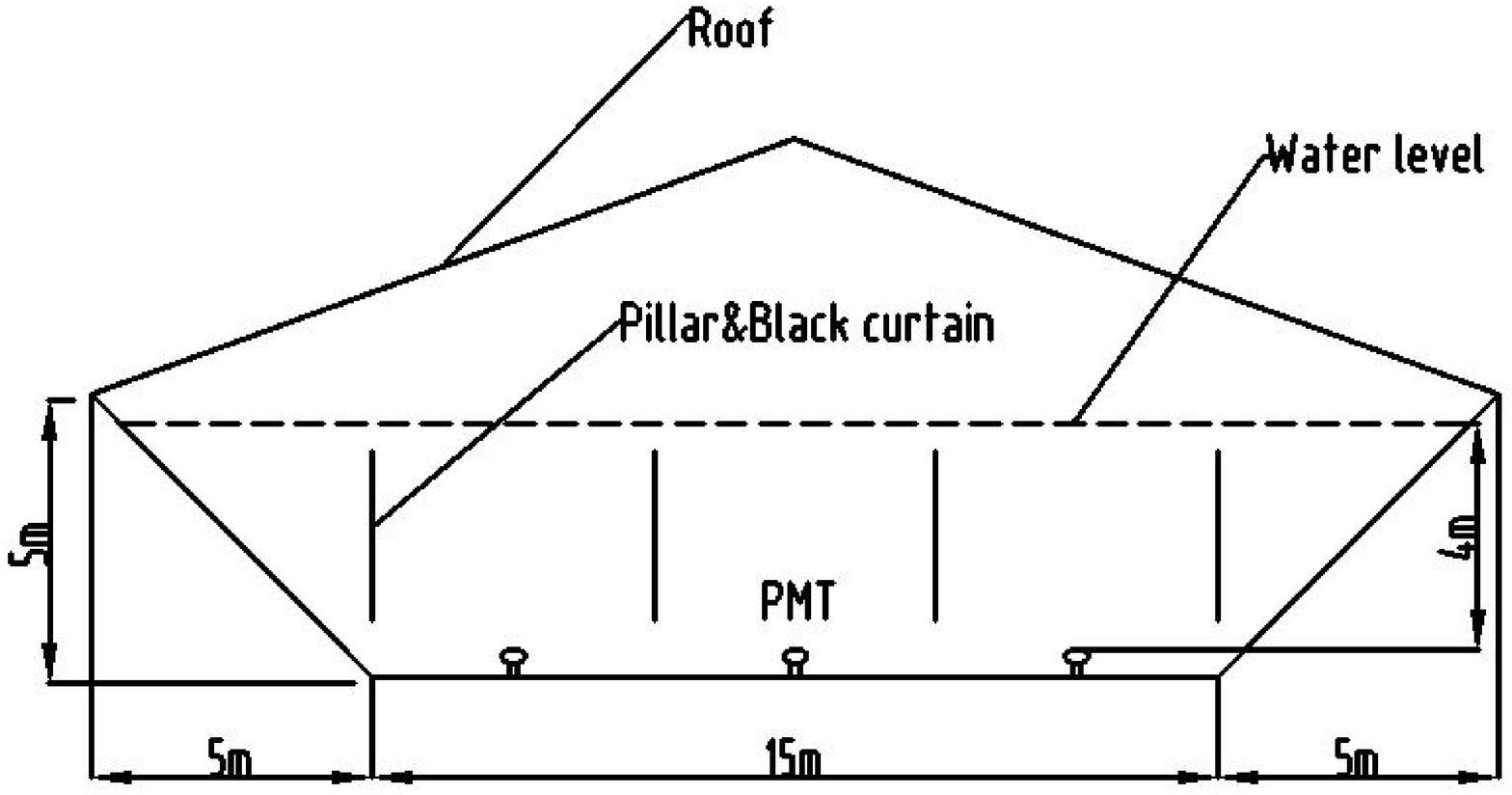}
  \figcaption{The tank of the prototype array. The four vertical lines are pillars made of ferroconcrete, and the black curtains hang from these pillars.}
  \label{fig:pond}
\end{center}

\subsection{Water purification and recirculating system}

To purify and keep the water clean, a water purification and recirculating system was designed and deployed in the control room~\cite{an2013}. This system consists of the following components: (1) a multi-media filter, (2) a carbon filter, (3) a fine filtration of 5~{\textmu}m, (4) a storage tank, (5) a fine filtration of 1~{\textmu}m, (6) an ultra-fine filtration of 0.22~{\textmu}m, and (7) a sterilization setup with UV lamps that have wavelengths of 254~nm and 185~nm. The UV lamp at 185~nm wavelength is a critical component because it can decompose the dissolved organic carbon, which is the major pollution source in the water. The original water pumped from a well is filled into the tank, and first passes through all of the above devices. Once the tank is full, the recirculating system starts to work. It pumps water from the top of the tank into a storage tank, then pumps the water from the storage tank back into the main tank using the last three devices mentioned above. In total, 1,600 tons of fresh filtered water is injected. The whole system has a filling capacity of 70~L/min. The circulation speed of this system is approximately 22 days per tank volume.

\subsection{Measurement of the water transparency}

To monitor the real-time water transparency, a tube device was designed and installed in the control room. The schematic of the device is exactly same as the one shown in Fig.~\ref{fig:tubedevice} in Section~1. The operation of the device is controlled by a slow control system, which supplies the power to the LED and PMT, and records the PMT signal and water level. This device can automatically and remotely measure the attenuation length of different water samples either several times a day or as required. Figure~\ref{fig:wattexample} shows a result of a measurement of the charge of PMT signals as a function of water level.

\begin{center}
  \centering \includegraphics[width=0.7\linewidth]{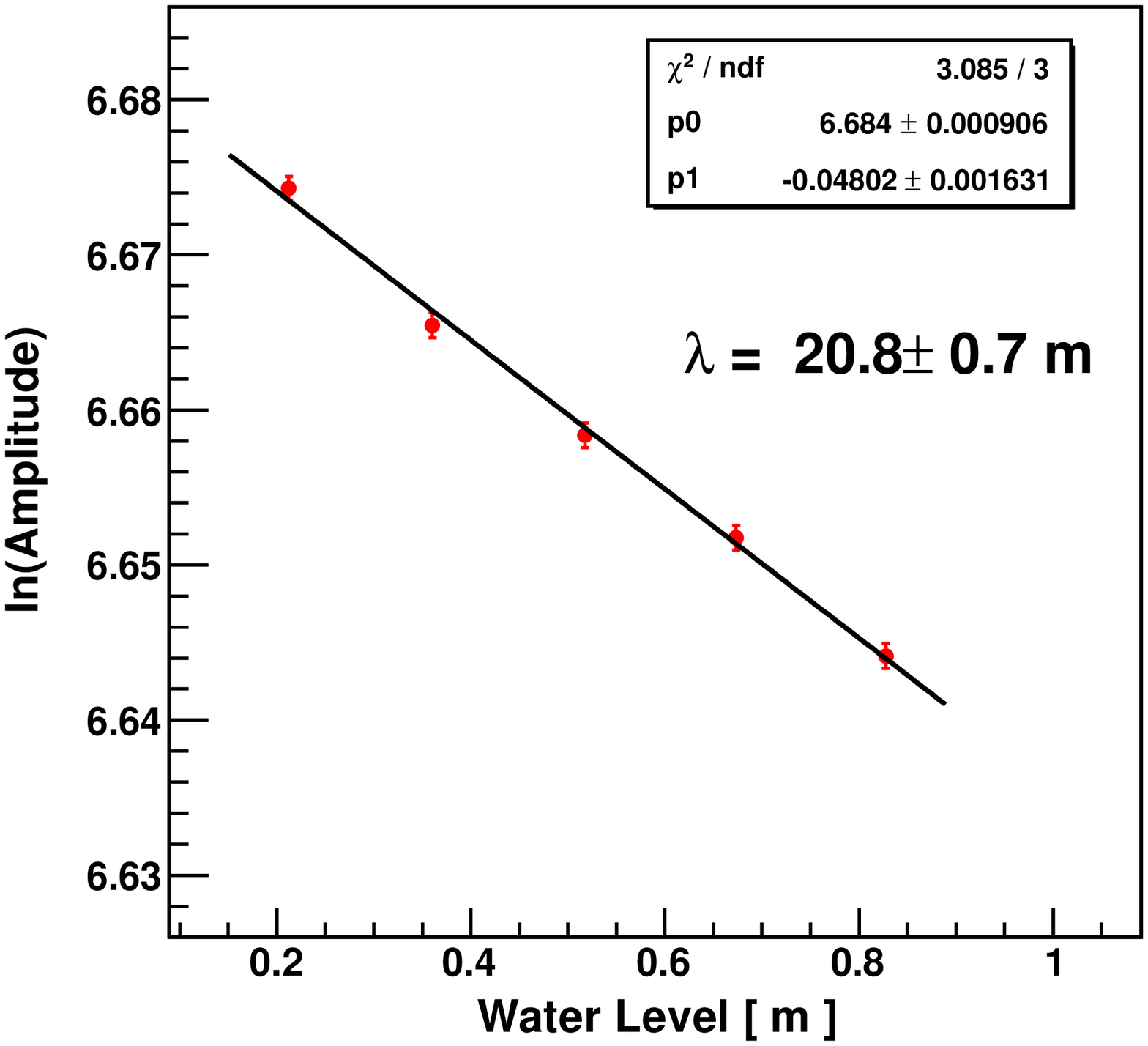}
  \figcaption{Fitting the signal intensity as a function of the water     depth in the tube. An exponential relationship $A=A_0\exp(-d/\lambda)$ is assumed, where $A$ is the average intensity of the LED pulsed signals measured by the PMT at every water depth $d$ in the tube, and $\lambda$ is the attenuation length.}
  \label{fig:wattexample}
\end{center}

In order to alleviate the manpower burden of  maintenance, the operation of the water purification and recirculating system of the prototype array has been paused from the end of 2012 to the middle of May, 2013. During this long period, the water in the tank became turbid, and the attenuation length was reduced to less than 4~m by the beginning of May. From May 15th 2013, the purification and recirculating system started to resume operation. Since the water was seriously polluted by bacteria and organic carbon, which could not be easily cleared away by the recirculating system, a chlorine-based chemical was added to the water in the tank so as to quickly kill the bacteria and disintegrate part of the organic compounds. With these enforcement measures, in addition to the continuous purification and recirculating process afterwards, the water transparency improved rapidly, and the attenuation length increased to more than 20~m after a month. Figure~\ref{fig:watt} shows the improving process of the water transparency, represented by the attenuation length at a wavelength of 405~nm, where the water samples drawn from the tank via the pipe were measured around 12 times per day. In the plot, only the data points averaged for around a day are shown, in order to weaken the temperature effect of the LED of the tube device during different hours of the day.

\begin{center}
  \centering \includegraphics[width=0.8\linewidth]{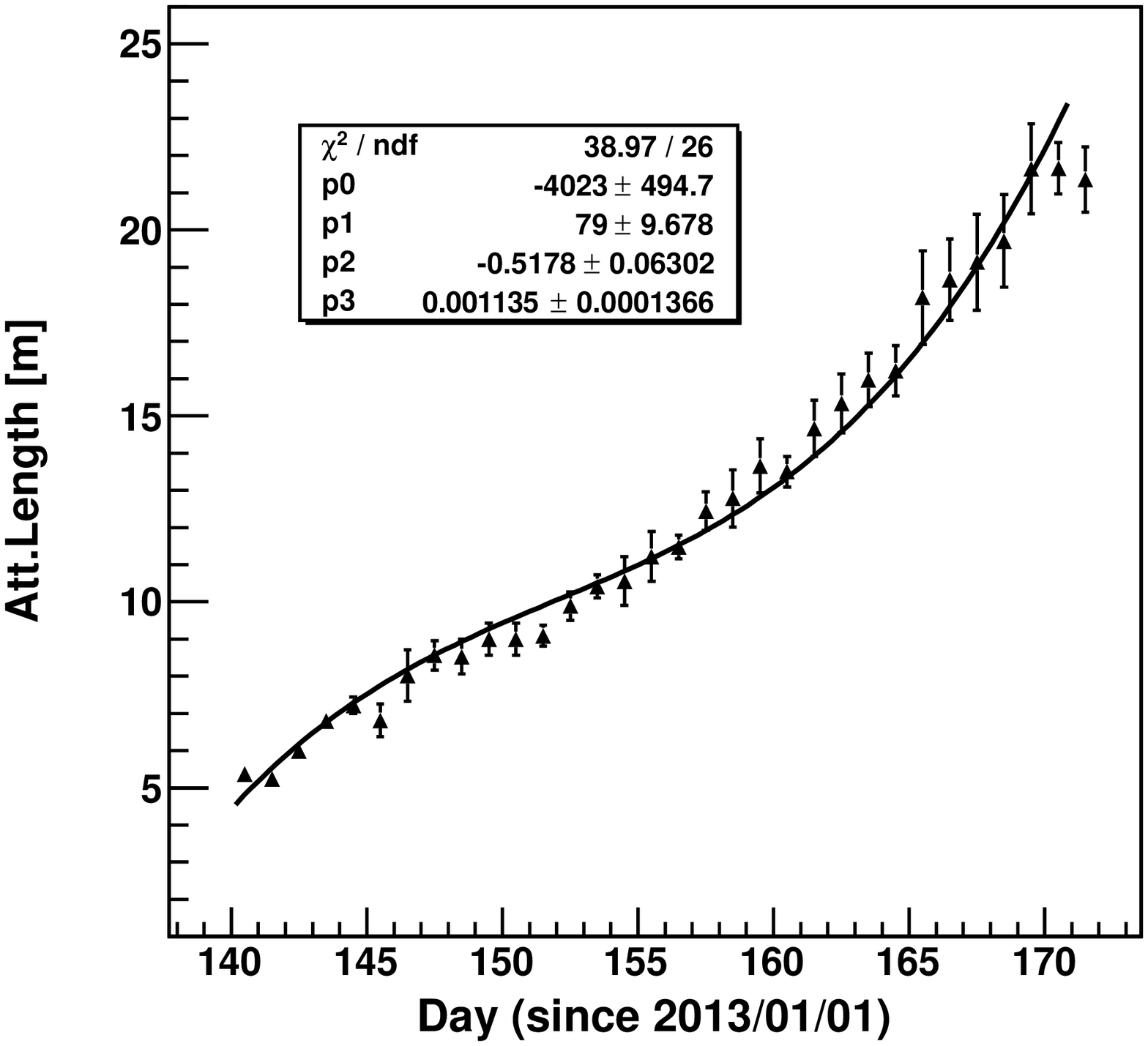}
  \figcaption{The variation of the water attenuation length during the period from May to June, 2013, where the solid line is a fit with a polynomial function. The samples reflect the average transparency of the water in the tank, and were obtained via the recirculating pipes.}
  \label{fig:watt}
\end{center}

Thanks to the water transparency being greatly increased during this period and a big effort being made in the measurement of the attenuation length with the tube device, the experimental data deserves to be analysed carefully, and a solution for water transparency monitoring and measurement with natural cosmic ray signals could be found.

\section{Single-channel signal}\label{sect:singlechannel}

\subsection{Data taking}
The data taking of the prototype array is specially designed to be multi-purpose, to allow different detector performance studies. Several data-taking modes (or namely, trigger modes) were performed repeatedly in turn, with the following settings for the period from May to June 2013 with which this study is concerned:
1) Single-channel signals with a low threshold (around 1/3 PE, where PE means photo-electron). These were taken approximately 8 times per day for every individual PMT channel (while other channels were masked), each lasting around 20 seconds;
2) Signals of any one of the channels passing with a high threshold (around 20 PEs). These were taken approximately 2 times per day and each lasted around 30 minutes;
3) Shower mode, which requires at least 3 PMTs firing at a low threshold during any 100~ns time window;
4) Other custom modes for particular analyses such as checking the electronics and testing the time calibration system.
Of these data-taking modes, only the first two are related to our study here. Combined with the data of these two modes, the single-channel charge distribution with a wide dynamic range for each PMT can be exactly derived, thus here we assign them a unified name, {\sl the single-channel mode}.

\subsection{Charge distribution of the single-channel mode}

In the single-channel mode, three peaks in the charge distribution are clearly observed, as shown in Fig.~\ref{fig:chargedis}. Previous analyses~\cite{yao2011,an2011} have proved that the first peak in the distribution results from the single photo-electron signals, and the third peak originates from nearly vertical cosmic muons directly hitting the photo-cathode. Two curves with the attenuation lengths around 22~m and 6~m respectively, are drawn in the same plot. Comparing two sets of curves, only the second peak position changes along with the water attenuation length. In other words, the position of the second peak is related to the water transparency.

Based on the Monte Carlo simulation, the second peak, with an amplitude around 10 PEs, has three sources. The major source is  pure muon signal, where its direction and position are smeared. Another is the electro-magnetic component of the showers, which usually produces Cherenkov light in the top tens of centimeters of the tank. The energy spectrum of the EM component  approximately obeys the power-law function. The simulation also shows that the hadronic component of the shower also has a small chance of forming the second peak, as its Cherenkov photon production behavior is similar to that of the electro-magnetic component. The charge distribution of signals originating from pure cosmic muons by MC simulation is also drawn in Fig.~\ref{fig:chargedis}, with the assumption of water attenuation length at different values. Obviously, all the three peaks appear in this charge distribution of cosmic muons, and the second one stands out with almost the  highest statistics. More analyses were done to explore the mechanism. Figure~\ref{fig:ddist} shows the distribution of the distance $D$, which is defined as the distance between the PMT and the point on the muon track where the Cherenkov photons are emitted, i.e., the travel distance of the Cherenkov photons. The most probable value is 3.0~m, in other words, photons caught by the PMT are generated at the same distance no matter how transparent the water is. Then, more photons are expected to reach to the PMT if the water is more transparent, as shown in Fig.~\ref{fig:qvsd}. When the corresponding distance is near to 3.0~m, the amplitude is around 10 PE, which corresponds to the position of the second peak shown in Fig.~\ref{fig:chargedis}. Thus, it can be concluded that the pure geometrical effect of the cosmic muons forms the second peak.

\begin{center}
  \centering \includegraphics[width=0.8\linewidth]{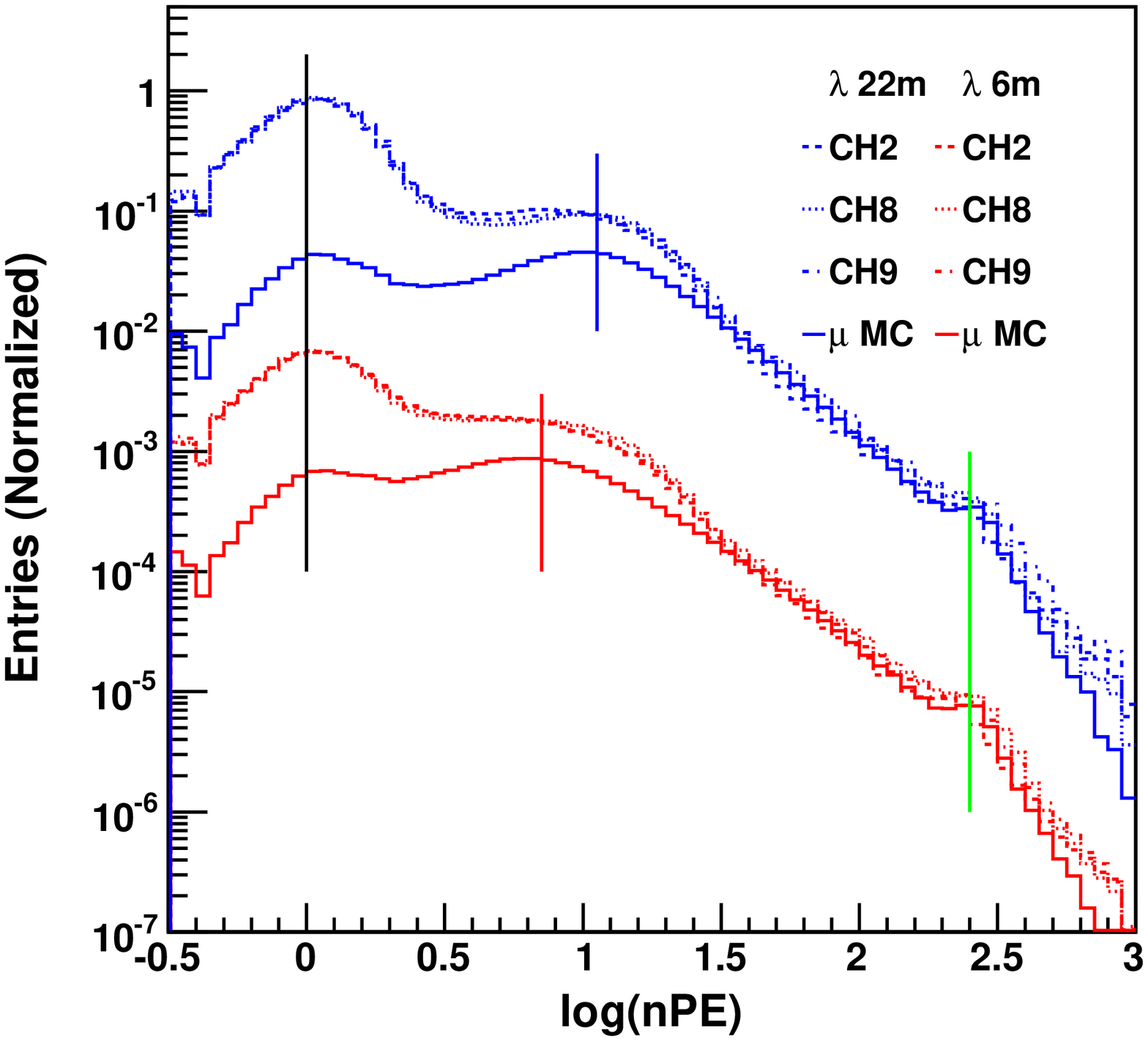}
  \figcaption{The charge distribution of several PMT signals in the single-channel mode (dotted line) and simulations of pure muons (solid line). The blue is for the attenuation length of 22 m (scales: 10 and 1.5); red is 6 m (scales: 0.1 and 0.02).  The  black  vertical line is the first peak; the green vertical line is the third peak; and only the second peak position changes along with the water attenuation length from red vertical line to blue vertical line. }
  \label{fig:chargedis}
\end{center}

In the simulation, a parameterized function for muon direction and momentum from Ref.~\cite{Bugaev1998}, with the flux measured by the vertical muon spectrum from CAPRICE~\cite{Kremer1999}, is used to sample the cosmic muons and applied to the generator of particles for the detector simulation.
Geant4 version 9.1.p01~\cite{Agostinelli2003} is used for tracking the muons and their productions in the detector, with the PMT models taken from the GenericLAND software library~\cite{neutrino}. The water absorption length at 405~nm is used to represent the water transparency. Values for other wavelengths are extrapolated from the curve for pure water measured by Ref.~\cite{Querry1}. The coefficient is scaled to 0.05, and the attenuation length is 20 m at the specified wavelength of 405~nm (Fig.~\ref{fig:absmodel}). Investigations of how the light is absorbed and scattered in pure water  and calculations of how the light is scattered by particulates in the water show that the attenuation length mainly leads to the absorption of visible light when the water transparency is not so good (e.g., an attenuation length less than 40~m). That means that the attenuation length is approximately equivalent to the absorption length here.

As mentioned above, the second peak is mainly caused by the geometrical effect of the muon tracks and the peak position depends strongly on the water transparency. It is reasonable to consider to extend this phenomenon in the data, so that finally a quick and efficient solution to measure the attenuation length of light in water with natural single cosmic muon signals could be achieved.

\begin{center}
  \setlength{\abovecaptionskip}{0pt}
  \setlength{\belowcaptionskip}{0pt}
  \centering
  \includegraphics[width=0.8\linewidth]{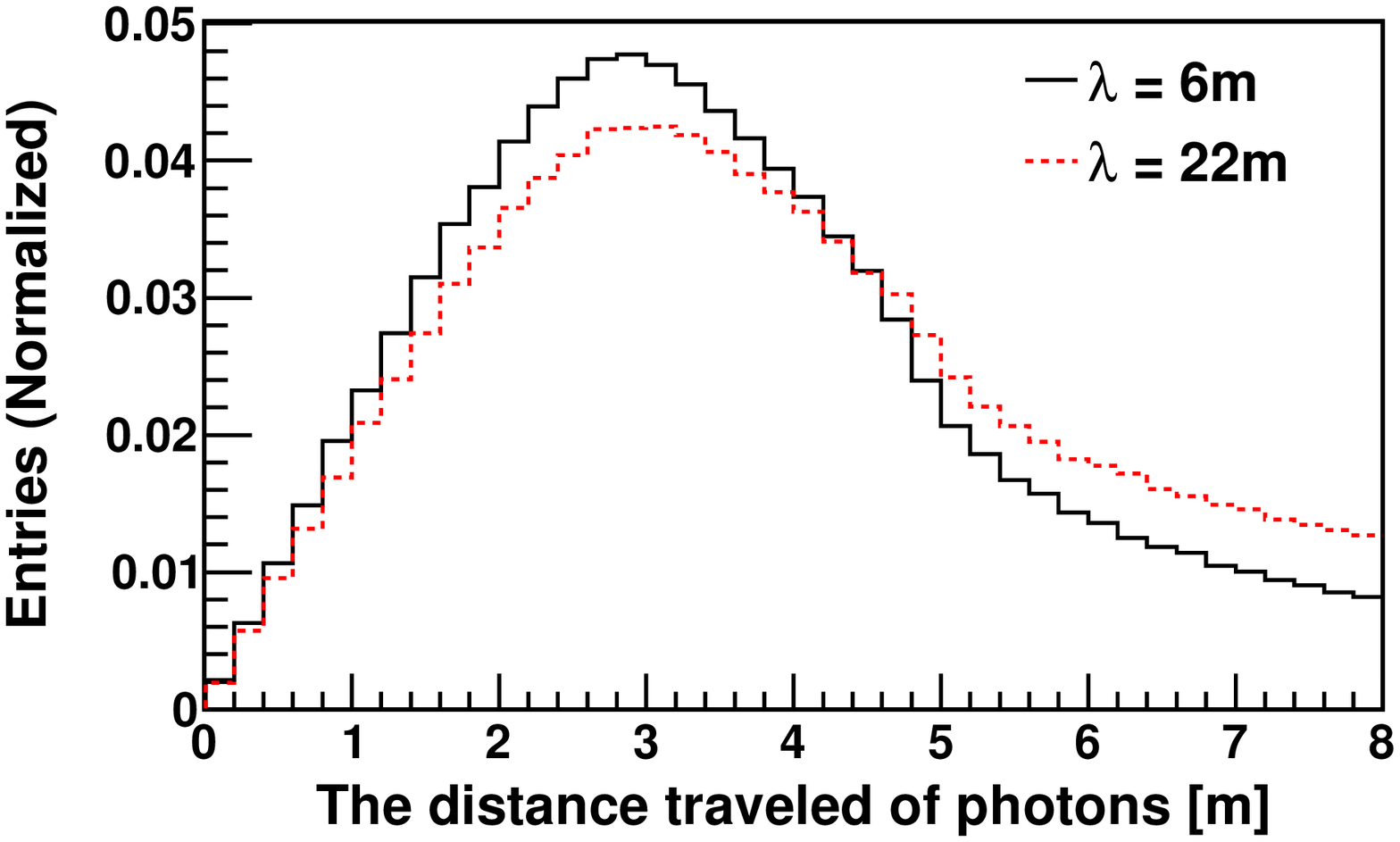}
  \figcaption{Distribution of the travel distance of the Cherenkov light. }
  \label{fig:ddist}
\end{center}

\begin{center}
  \setlength{\abovecaptionskip}{0pt}
  \setlength{\belowcaptionskip}{0pt}
  \centering \includegraphics[width=0.8\linewidth]{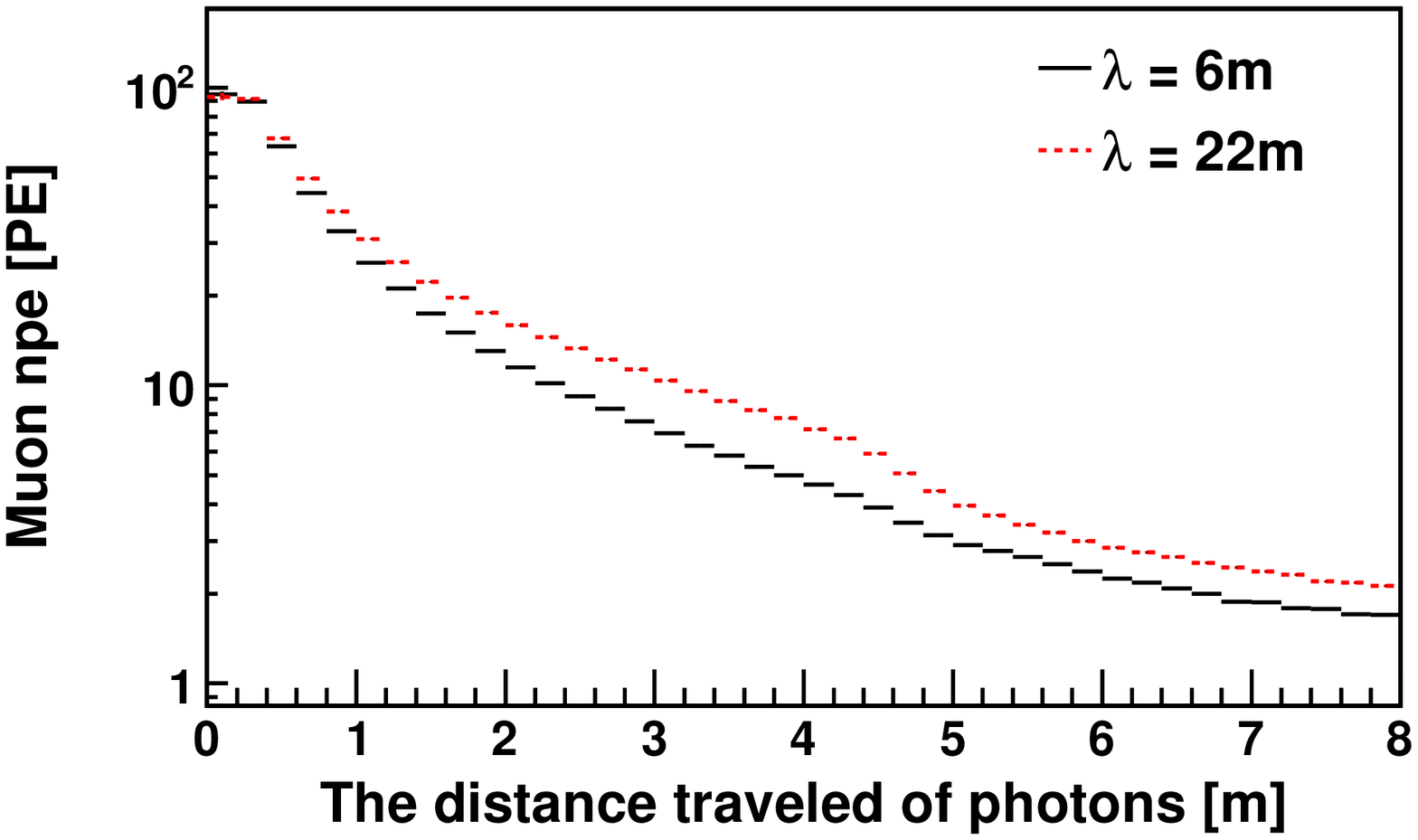}
  \figcaption{Correlation between the signal charge and the travel distance     of the Cherenkov photons produced by cosmic muons.}
  \label{fig:qvsd}
\end{center}

\begin{center}
  \setlength{\abovecaptionskip}{0pt}
  \setlength{\belowcaptionskip}{0pt}
  \centering
  \includegraphics[width=0.8\linewidth,clip]{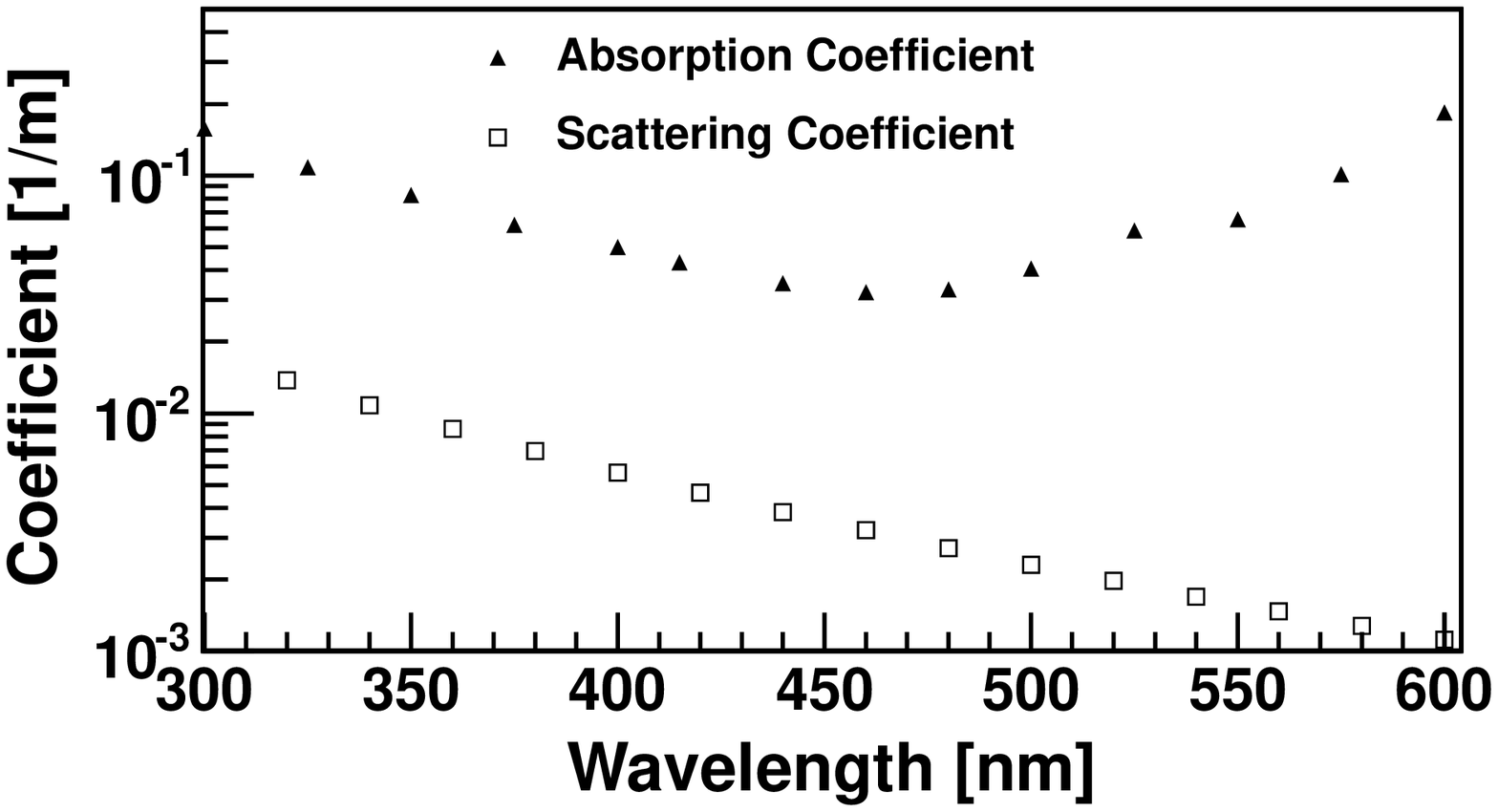}
  \figcaption{The absorption and scattering coefficient (reciprocal of
    the length) of the water transparency adopted in the simulation.}
  \label{fig:absmodel}
\end{center}

\section{Analysis of the second peak for the attenuation length measurement}\label{sect:analysis}

\subsection{Analysis of the peak position}

The charge distribution of single-channel signals is related to many random nature sources, such as the shower primary and its energy distribution, the distribution and the fluctuation of the type, the number and the energy of the shower particles falling into the detector cell, the distribution and fluctuation of number of Cherenkov photons generated and arrived at the PMT photo-cathode, the amount and the fluctuation of the charge value measured by the PMT and electronics, the noise generated by the PMT and the environment (e.g., radon radioactivity in the water), and so on.  Of the above sources, the most complicated source is the muon signals, as it depends strongly on  geometrical factors such as the incident position and the direction. Generally, it is not easily described by a simple distribution curve. Therefore, it is almost impossible to fit the charge distribution of the single-channel signals in a wide range by a simple function with only a few parameters.

To solve this problem, a quick but practical method is to fit only the neighboring range around the second peak which we are interested in, after a simple scale to the curve. Considering that the energy distribution of the cosmic rays and their secondaries approaches very much a power law function, a basic idea is to multiply a power law of the charge to the entries to obtain a rather flat curve, so that peak shape can be extruded. Actually in  Fig.~\ref{fig:chargedis}, such a power-law multiplication has been applied already, as the binning of the charge is logarithmic in this analysis, which is equivalent to a $Q$ ($Q$ is the charge) factor multiplication. Based on this idea, the curve is finally multplied by a power law function $Q^{1.5}$ with a linear binning. As shown in Fig.~\ref{fig:powerlaw}, it seems that the obtained curve turns quite asymmetrical around the peak, then a Gaussian function can be employed to fit the curve. We also tried with other power law indexes, but not much difference was observed. A similar way has been applied to the third peak~\cite{gao2014} for obtaining the peak position (power law index: 2.5) that will be used as a charge calibration parameter.

\begin{center}
  \centering \includegraphics[width=0.9\linewidth]{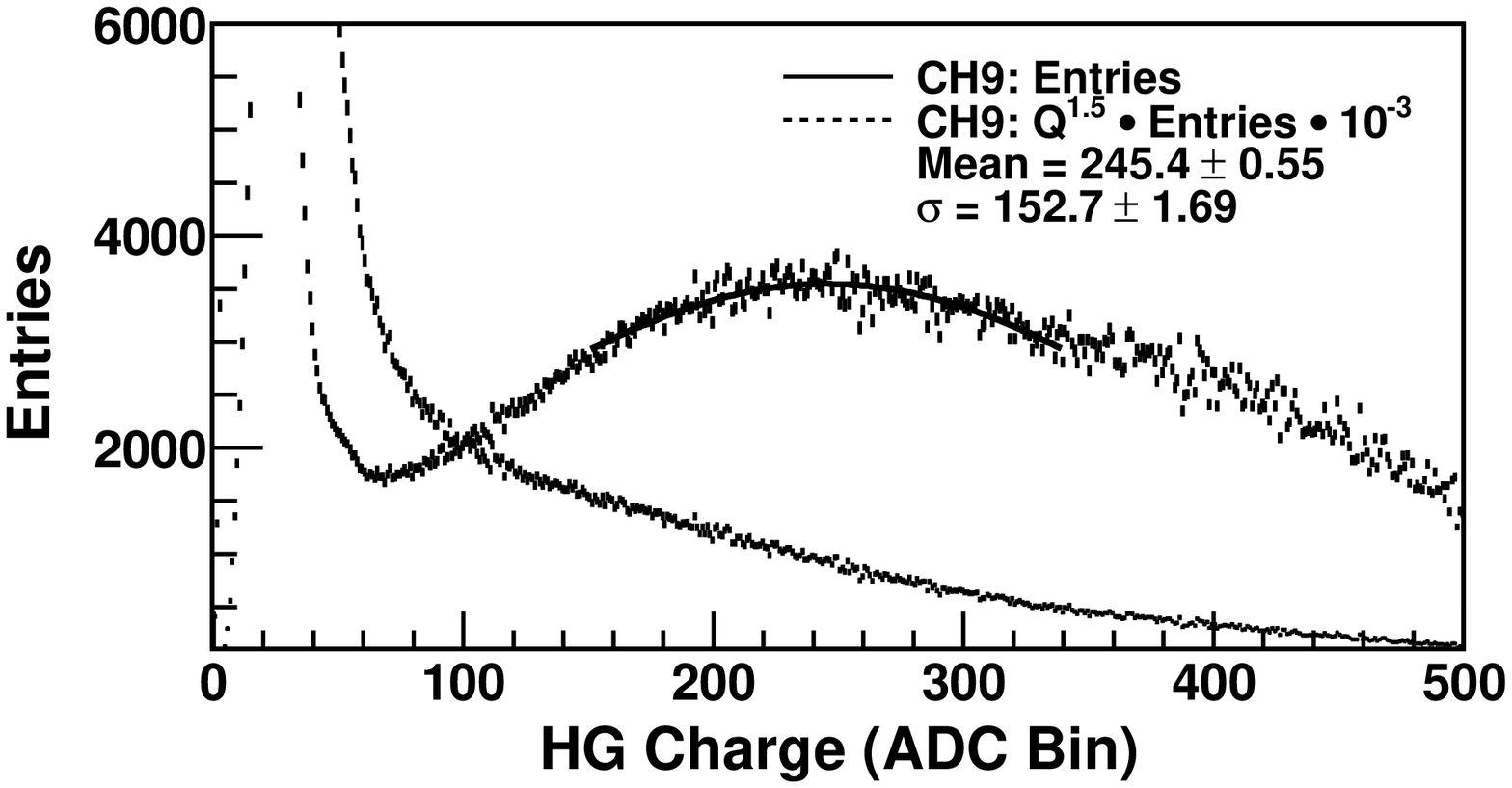}
  \figcaption{The second peak of a PMT from the prototype array (channel 9) before and after a power law multiplication (power law index: 1.5), where $Q$ is the high gain charge in number of ADC counts.  The curve with the power law multiplication is fitted with a Gaussian function, and the peak position is then obtained.}
  \label{fig:powerlaw}
\end{center}

We also analyzed the consequence on how the power law multiplication influences the systematic uncertainty of the peak position. It shows that a scale factor imposed to the charge will linearly shift the peak position, therefore, this method is quite suitable and the systematic error of the charge measurement will not be distorted in this procedure.

\subsection{Analysis of the efficiency difference}

As mentioned in Section~2, a layer of dirty coating had adhered to each PMT surface due to the long time deterioration of the water in the tank. The layer of coating will absorb some part of the Cherenkov photons, and reduce the detected light intensity. This reduction, called the surface efficiency here, could be regarded as the fourth factor besides the nominal quantum efficiency, the collection efficiency of the photo-electrons in the first dynode, and collection efficiency of electrons in the rest of the dynodes until the anode. The total efficiency of PMT is made up of the above four factors.

In order to smooth the efficiency difference of the PMTs, the third peak in the charge
distribution is adopted. This method is as follows. Firstly, the value of the third peak was obtained for every PMT channel (A), then the average value (B) of all available channels was calculated. With regard to an individual PMT channel, a scaling factor of average value B to value A is obtained, which reflects the relative efficiency for this PMT.

During this period, among all 9 cells of the prototype array, 6 had the same type of PMTs and were deployed with the same geometrical configuration. One cell was surrounded with Tyvek film for muon detection studies, and of the  other 5 cells, only 3 PMTs operated well throughout the measurement period. So only these 3 PMTs data are analyzed here and the scaling factors of these 3 channels are listed in Table~\ref{peak-iii}. We have checked the data with  different water transparencies during the same period, and the results show that the difference of the scaling factor is less than 1\%.

\begin{center}
  \tabcaption{The values of third peak and the obtained scaling factors. The attenuation length is around 20~m for the data used for this analysis.}
  \label{peak-iii}
  \begin{tabular}{lcccc}
    \hline
    PMT channel & CH2 & CH8 & CH9 & Mean\\
    Third peak (PE) &260&296&286&281\\
    Scaling factor &0.93&1.05&1.02&--\\
    \hline
  \end{tabular}
\vspace{0mm}
\end{center}
\vspace{0mm}

\subsection{Peak position versus attenuation length}

As mentioned in Section~2, there are several kinds of data-taking modes, operated repeatedly. The single-channel data with the low threshold can be used for the second peak analysis, and around 8 measurements were taken every day. The data was not continuously taken during the whole period because sometimes it was occupied by a dedicated trigger mode test for the time calibration system. Nevertheless, it is not a serious problem for the analysis. The attenuation length measured directly  with the tube device as shown in Fig.~\ref{fig:watt} in Section~2 was applied, and the value of the attenuation length at any moment can be obtained by evaluating the fitted polynomial function. The position of the second peak for different PMT channels as a function of the attenuation length is given in Fig.~\ref{fig:peakvswatt}, and the data points for each channel are fitted by an exponential function as the following Equation~\ref{eq:eq1}:

\begin{eqnarray}
  N_{\rm 1} = N_{0} \cdot e^{(-d/\lambda)}.
  \label{eq:eq1}
\end{eqnarray}

where $N_{\rm 1}$ is the second peak position, $N_0$ is the peak position when the transparency is infinite, $d$ is the most probable value of travel distance of the Cherenkov photons, and $\lambda$ is the attenuation length.
The relative average deviation of different PMT channels is 3.1\%.
The same procedure was  applied to the data from 2012, when the water transparency was always good and the PMT surface was very clean. The result shows that the difference in the PMTs is around 1.5\%.
In the course of water injection or some accidents, the water attenuation length can change between 6~m  and 20~m, so the variation range of the second peak position is nearly 35\%. Therefore, the difference of PMTs  has little influence on accurately measuring and monitoring the water transparency  and its variation.

\begin{center}
  \centering \includegraphics[width=0.9\linewidth]{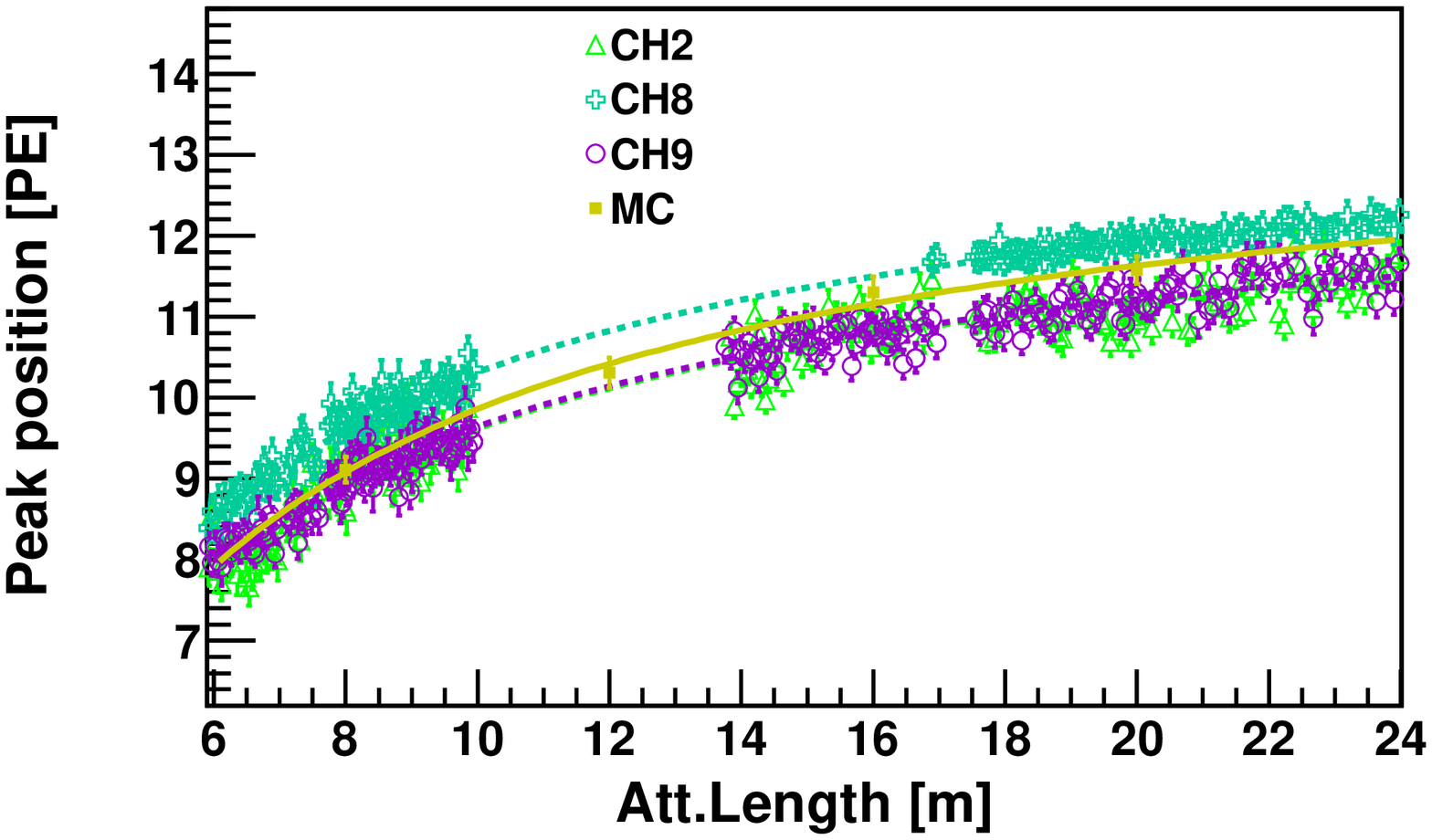}
  \figcaption{The second peak position as a function of the attenuation length. Different markers denote different PMT channels and the simulation data, marked with ``MC'', as shown by the text in the plot. For comparison, the data points for each channel and MC are shown, with fits by Equation ~\ref{eq:eq1}.}
  \label{fig:peakvswatt}
\end{center}

For comparison, a Monte Carlo simulation curve is also drawn in Fig.~\ref{fig:peakvswatt}. In the simulation, the air showers events are generated by Corsika v75000~\cite{corsika} and the QGSJET-II model~\cite{Ostapchenko2006} is used for high energy hadronic interactions. The injection area of the shower cores is $\rm 20\;km\times 20\;km$, and the sampled energy for the primary energy is in the range from several GeV to 1 TeV, using fluxes measured by AMS-02~\cite{Aguilar1,Aguilar2} and CREAM-II~\cite{Ahn2009}. The detector simulation is the same as that mentioned in Section~3.
The simulation curve fits well to that of the data, implying that the relationship between the peak position and the attenuation length of water is understandable in terms of the origin of cosmic muons among the shower particles.

\subsection{Error analysis}

Since the main purpose of our study focuses on monitoring and measuring the transparency of the water with analysis of the second peak position using the single-channel signals, the relation between the error of the peak position and the measurement error of the attenuation length is discussed here. Conversion between these two kinds of errors can be achieved with a transformation of Equation~\ref{eq:eq1}, i.e.,

\begin{eqnarray}
  \frac{\Delta\lambda}{\lambda} = \frac{\Delta N_{\rm 1}}{N_{\rm 1}}
  \frac{\lambda}{d}.
  \label{eq:eq2}
\end{eqnarray}

As for the 2nd peak position analysis, for each data point, 20 seconds of single-channel data are used.  The error of peak position at a particular water transparency in principle could be given by the well-developed fitting procedure, i.e., the error of the fitted parameter; however, this is too optimistic as the error is very small. A more practical and reasonable method is to calculate the fluctuations of a couple of the peak positions for a day, as the water transparency changes very little in such a short period. The RMS of this fluctuation for each PMT is evaluated, representing the error. Converting the error into a relative value, and averaging all the available channels, the error of the attenuation length is finally obtained with Equation~\ref{eq:eq2}.
The error is  controlled at the 6\% level even at a very good water transparency, for instance 22~m,  and is actually a little better than the precision around 7\% measured with the tube device at the similar water transparency.

\section{Conclusions}\label{sect:Conclusion}

Natural sources such as cosmic muons can produce a three-peak feature in the charge distribution of the PMT single-channel signals in our experiment. By analyzing the second peak position in the distribution, the attenuation length of the water can be monitored and measured with a precision of 6\%.

In the LHAASO-WCDA experiment, the single-channel signals produced by each PMT will be read out via a dynode and anode in order to gain a wider dynamic range, and a triggerless mechanism for the data-taking will be adopted. In the triggerless mechanism, the single-channel data of every PMT channel will be transferred to a computer cluster for further processing, such as filling histograms, forming triggers, and even going through an online reconstruction for filtering out noise.  During this process, the charge of the single-channel signals for every PMT will be filled into histograms, stored once into buffer every few tens of minutes. This enables a real-time analysis of the second peak more than 50 times per day with much better control of the fluctuations, as the statistics are huge. That means the transparency of the water can be measured precisely in a continuous and exhaustive way, with a granularity up to a detector cell. This method provides a better alternative to the function of the dedicated tube devices. The analysis procedure developed in this study will further ensure the detector-unit charge signal uniformity, and the energy can be better measured.

\section*{Acknowledgments}
The authors would like to express their gratitude to X.~F.~Yuan, G.~Yang, W.~Y.~Chen and C.~Y.~Zhao for their essential support during the installation, commissioning and maintenance of the prototype array.

\end{multicols}
\clearpage
\end{CJK*}

\end{document}